\documentclass[runningheads]{llncs}
\usepackage{textcomp}
\usepackage{graphicx}
\usepackage{xcolor}
\newcommand\blfootnote[1]{%
  \begingroup
  \renewcommand\thefootnote{}\footnote{#1}%
  \addtocounter{footnote}{-1}%
  \endgroup
}

\begin{document}
\title{Colorectal Cancer Segmentation using Atrous Convolution and Residual Enhanced UNet}
\author{Nisarg A. Shah\inst{1} \and
Divij Gupta\inst{1} \and
Romil Lodaya\inst{2} \and Ujjwal Baid\inst{2} \and Sanjay Talbar\inst{2}}
\authorrunning{N. Shah et al.}
%
\institute{Department of Electrical Engineering, Indian Institute of Technology Jodhpur, India\\ \email{\{shah.2,gupta.13\}@iitj.ac.in} \and
Shri Guruji Gobind Singhji Institute of Engineering and Technology, Nanded, India\\
\email{\{romillodaya3007,ujjwalbaid0408\}@gmail.com}, \email{sntalbar@sggs.ac.in}}

\maketitle              
%
\blfootnote{D. Gupta, R. Lodaya, U. Baid contributed equally to this article}
\begin{abstract}
Colorectal cancer is a leading cause of death worldwide. However, early diagnosis dramatically increases the chances of survival, for which it is crucial to identify the tumor in the body. Since its imaging uses high-resolution techniques, annotating the tumor is time-consuming and requires particular expertise. Lately, methods built upon Convolutional Neural Networks(CNNs) have proven to be at par, if not better in many biomedical segmentation tasks. For the task at hand, we propose another CNN-based approach, which uses atrous convolutions and residual connections besides the conventional filters. The training and inference were made using an efficient patch-based approach, which significantly reduced unnecessary computations. The proposed AtResUNet was trained on the DigestPath 2019 Challenge dataset for colorectal cancer segmentation with results having a Dice Coefficient of 0.748. 
Its ensemble, with its simpler version, achieved a Dice Coefficient of 0.753. 
  
\end{abstract}

\section{Introduction}

Cancer is the abnormal growth of cells that can invade or spread to other parts of the body. Colorectal cancer is a type of cancer that begins in the large intestine (colon). This cancer is often seen in old age people, but now it can even be seen in younger people due to lifestyle factors, with only a small number of cases due to underlying genetic disorders. Colorectal cancer is the fourth most common cancer diagnosed and the third leading cause of cancer death worldwide  \cite{surveycancer}. Chances of survival increase manifold if the cancer is diagnosed early. This cancer is diagnosed by obtaining tissue samples by Colonoscopy. These tissues are stained using hematoxylin and eosin(H\&E) stain. The hematoxylin stains the cell nuclei blue, and eosin stains the extracellular matrix and cytoplasm pink, with other structures taking on different shades, hues, and combinations of these colors. The glass slides which contain the stained tissue are digitized and converted into high-resolution whole slide images(WSI). Their diagnosis requires experienced pathologists and is also a laborious task. Since the images are high-resolution, it is a challenging task to make an automatic segmentation tool that accurately predicts the tumor region. However, recently, Deep Learning(DL) approaches have shown to be much better than the conventional techniques for segmentation tasks, with many researchers worldwide publishing various work on the same. In this paper, we propose another Convolutional Neural Network(CNN)-based DL approach for the segmentation of the tumor wherein we used a patch-based, sliding window technique as the images used for the task were of high resolution. In this paper, we present a novel convolutional block based on the concept of atrous convolutions. The block can be easily integrated into other CNN-based approaches as well. We also make use of a simple pre-processing and post-processing approach for better results.


\section{Related Work}
The power of CNNs was first exhibited in the ImageNet challenge \cite{imagenet}, and ever since then, CNNs have revolutionized the field of computer vision and have produced far better results than conventional techniques on numerous tasks. The same can be said in the case of medical imaging\cite{raj2020multivariate}, particularly segmentation. The inherent property of CNNs to automatically find crucial and task-relevant structures in images account for their widespread use. The first revolutionary architecture was the UNet \cite{uneto} introduced for the task of cell tracking and segmentation. The UNet \cite{uneto,3dunet,baid2020novel} is a well known CNN architecture first introduced in 2015 primarily for cell segmentation. Since then, it has been the backbone of several biomedical segmentation architectures.
The UNet \cite{uneto} consists of a contracting path (encoder) and an expansion path (decoder), along with the skip connections in between the corresponding layers of the encoder-decoder to retain the spatial information between early and late layers for location precise segmentation maps. Many researchers have modified the UNet to produce impressive results on various biomedical segmentation tasks. In  \cite{baid2019brain}, the authors used a cascaded UNet for brain tumor segmentation, while in  \cite{raunet}, the authors used attention-mechanism in the UNet for liver tumor segmentation. In  \cite{bladder}, the author varied the kernel size of the filters in the UNet for bladder cancer cell segmentation. Researchers have recently shifted their focus on using deep learning techniques for histopathology analysis, especially colorectal cancer diagnosis. In  \cite{sirinukunwattana2016locality}, the authors have discussed the use of locality-sensitive deep learning with the use of Spatially Constrained CNN. In \cite{kather2019predicting}, the authors have discussed the prediction of the clinical course of patients diagnosed with colorectal cancer, while in  \cite{bychkov2018deep}, the authors have discussed estimating the patient risk score using LSTMs. Also, some work has also been done for incorporating adversarial or GAN-based approaches as in the work of  \cite{zhu2020multi} wherein the authors have also used concepts of attention, pyramid pooling, and atrous convolutions in their work. Another work by  \cite{figueira2020adversarial} uses adversarial approach for domain adaptation to detect the tumor in an unsupervised manner. Another popular approach is the use of ensembles such as in the work of \cite{khened2020generalized} wherein the final prediction was obtained after averaging predictions from three FCN models.

\section{Method}
This section discusses the proposed method, which primarily uses CNNs with the UNet as the backbone. We provide an in-depth discussion about the various components of the architecture and their features and, finally, the whole architecture as one.

\subsection{Atrous convolution}
Atrous convolution  \cite{atrous_conv} is a convolutional operation wherein an extra parameter, the dilation rate, is used in addition to the convolutional layer. The dilation rate determines the spacing between the values in a kernel. By dilating the convolutional kernel, a broader receptive field is acted upon for the same computational cost as that of a regular convolution operation. This type of convolution is particularly useful in segmentation, which requires feature extraction from various receptive fields. The atrous convolutions have been shown to decrease blurring in semantic segmentation maps. Additionally, they are indicated to produce the same effect as that of pooling by extracting long-range information. 

\begin{figure*}[t]
	\begin{center}
		\includegraphics[scale=0.95]{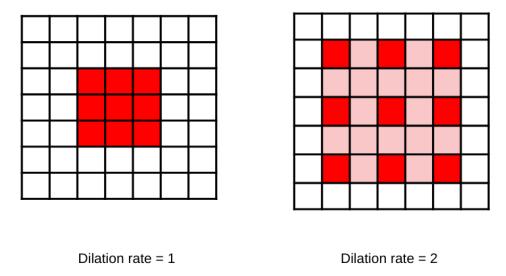}
		\caption{Effect of change of dilation rate. With increasing the dilation rate, the space between the weights(dark red) increases and is filled with zeros. In this manner, the receptive field is increased.}
		\label{patch}
	\end{center}
\end{figure*}
\subsection{Series Atrous Convolution Unit}
\begin{figure*}
  \includegraphics[width=\textwidth,height=2.5cm]{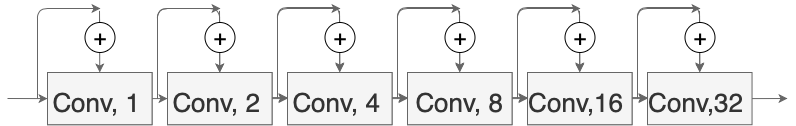}
  \caption{The Series Atrous Convolution Unit}
  \label{series_atrous_convolution}

\end{figure*}

\begin{figure*}
    
	\includegraphics[trim=4cm 0 0 0cm,width=16cm,height=8cm]{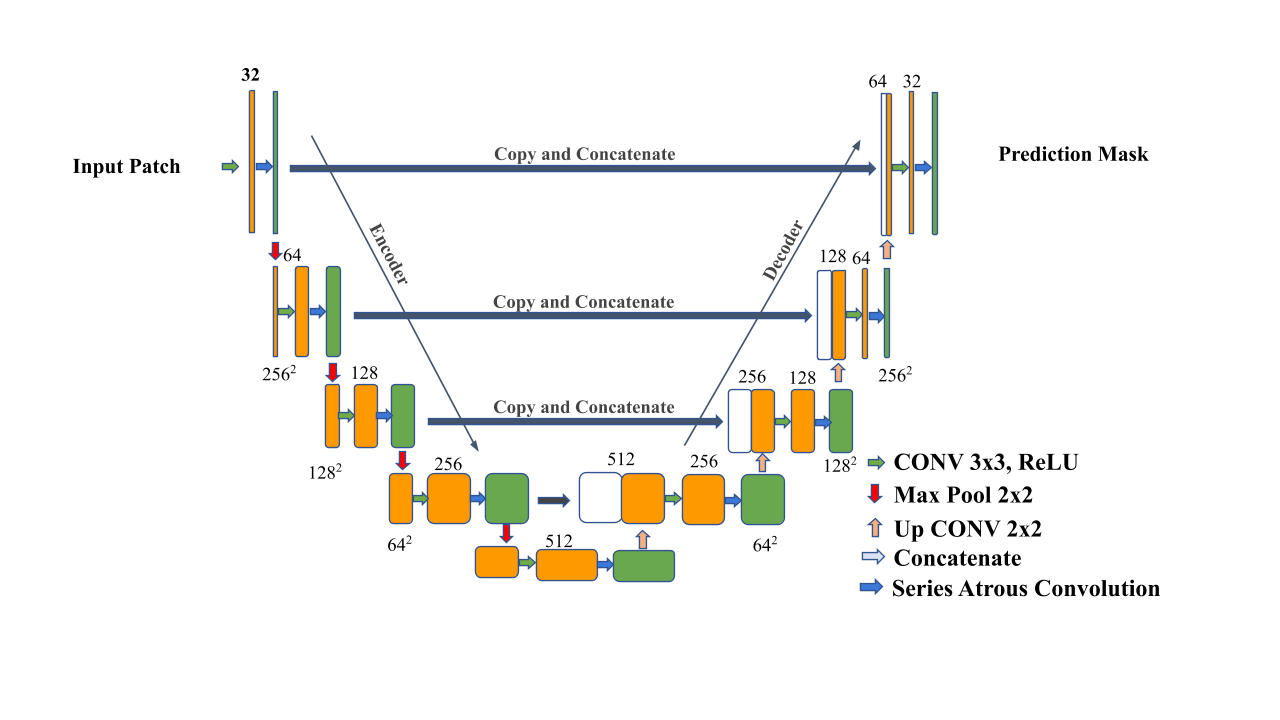}
	\caption{The proposed AtResUNet}
	\label{architecture}
	
\end{figure*}

The Series Atrous Convolution Unit makes use of series pixel-wise addition on the feature map obtained from a series of convolution operations done at a particular dilation rate, as shown in Figure \ref{architecture}. This is similar to using residual connections  \cite{resi_conn}. Using residual connections also ensured that information was not diminished, as in the general case of deep networks. Experimentally, we found that a series connection of feature maps obtained at different dilation rates produced better results than a concatenation of convolution operations at different dilation rates. We tried different types of combination for the series Atrous Convolution Unit like (1,2), (1,2,4), (1,2,4,8), (1,2,4,8,16), (1,2,4,8,16,32), and the best combination among these experiments was obtained for (1,2,4,8,16,32), based on the segmentation results. Due to computational limitations, we did not experiment with every possible combination of the dilation rates. Therefore, the particular combination (1,2,4,8,16,32) was used in all further experiments.
The Series Atrous Convolution Unit is shown in figure \ref{series_atrous_convolution}.

We represent the Series Atrous Convolution Unit as following :- 

\[F_i(x) = w\oplus_i x + b\] 
The above equation indicates the output \textit{F} for input \textit{x} after convolution with 3x3 kernel, \textit{w} with dilation \textit{i} and bias \textit{b}. For the proposed Series Atrous Convolution Unit, \textit{i} can take up values, 1,2,4,8,16,32.  

With the above terminology, we define the Series Atrous Convolution Unit as below.

\[Input = x_1\]

\[x_2 = F_1(x_1)+x_1\]

\[x_3 = F_2(x_2)+x_2\]

\[x_4 = F_4(x_3)+x_3\]

\[x_5 = F_8(x_4)+x_4\]

\[x_6 = F_{1_6}(x_5)+x_5\]

\[x_7 = F_{3_2}(x_6)+x_6\]

\[Output = x_7\]

\subsection{Proposed Architecture}
In the proposed architecture depicted in figure \ref{architecture}, the UNet \cite{uneto} is used as the base model, and the Series Atrous Convolution Unit is used for feature extraction. The input image is passed through a convolutional layer from which basic feature extraction occurs at that particular resolution. After that, the extracted primitive map is fed into the Series Atrous Convolution Unit having the same number of filters as the input feature map. 
The crucial information is extracted by $1\times1$ convolution, which decreases the number of channels in feature maps to that of the input to the Series Atrous Convolution Unit. The presence of Series Atrous Convolution Unit aids in extracting meaningful information required for the model training, as well as for proper convergence of the model. Moreover, the presence of the Series Atrous Convolution Units in the expansion region of the proposed model architecture helps in streamlining essential features present in the feature map obtained from the upsampling operation and enhances them considerably. Skip connections, inherent to UNet, share information at the same encoder-decoder level, which helps boost segmentation accuracy through the proper flow of gradient through the model. However, the apparent drawback to UNet-style architectures is that training of the intermediate layers of deeper models gets sluggish, which increases the risk of the network learning to scorn away the layers where abstract features are extracted. Conceivably, using a UNet architecture can improve the retention of fine detail features with fast training of deep networks. The benefits of the UNet outweigh its drawbacks, which prompted us to use it as our base model upon which we base our improvements.

\section{Data Processing and Training}
\subsection{Dataset}
The DigestPath, 2019  \cite{digestpath1} dataset was used which consisted of colonoscopy images of 750 tissue samples from 450 patients. The challenge also provided another dataset on signet ring cell detection. The average size of the images in the dataset was 3000x3000. This data was collected from multiple medical centers from several small centers in developing countries/regions; hence, it shows a significant appearance variation. Image style differences can be an obstacle for the screening task. The tissue samples collected were first dehydrated and then embedded in melted paraffin wax. After that, the resulting block was mounted on a microtome and cut into thin slices. All whole slide images were stained by hematoxylin and eosin(H\&E) and scanned at X20. We applied standard data augmentation techniques such as rotating, flipping, shear and stretch. The dataset was split into 75\% for training, 15\% for validation, and 10\% for testing.

%

\subsection{Preprocessing}
\label{ssec:subhead}
In the training phase, the model was trained with 50\% overlapping patches of size 512x512x3. This was primarily done to mitigate the effects of class imbalance and the less availability of data. The data consists of a white background and tissue sample in the foreground. Therefore, when patches of 512x512x3 were generated, many patches consisted of only the white background or very less useful portion. These redundant patches would have misled the training, so they were discarded. The samples' discarding was based on thresholding the amount of tissue sample or the useful information present in the patch. The dataset was segregated by thresholding it for a minimum X\% of tissue pixels. After thresholding for several percentages such as 40\%, 30\%, 20\%, etc., it was observed that by thresholding with 30\%, maximum redundancy was removed, and useful information was saved. Also, the patches with data less than 30\% would be covered in other patches that share the same 50\% overlap. Lastly, the patches were directly normalized to 0-1 by dividing each pixel by 255. This normalization has the effect of stabilizing the learning and converging of the model, requiring less training.
\begin{figure*}[t]
	\begin{center}
		\includegraphics[scale=0.52]{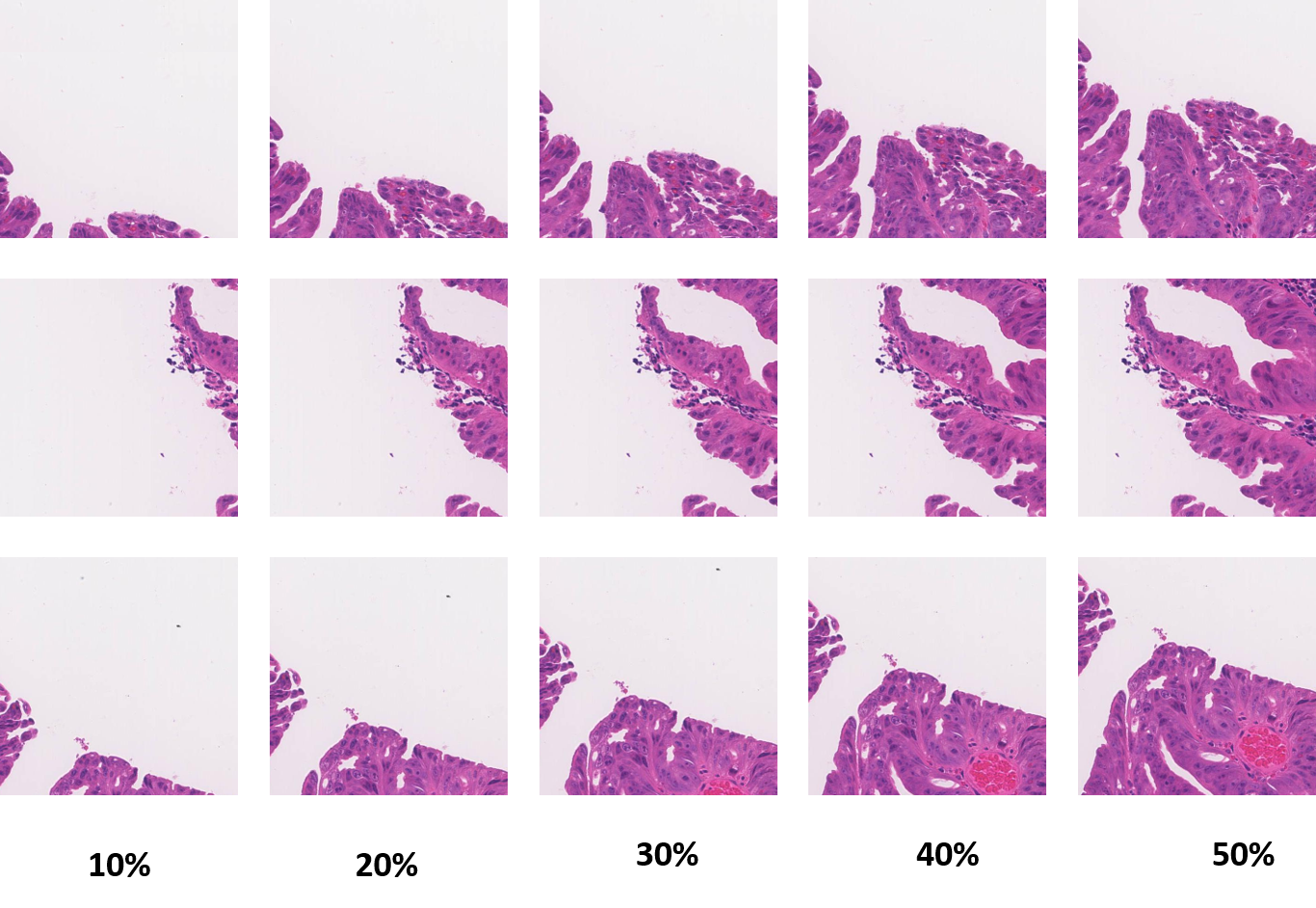}
		\caption{Sample patches extracted for training. \% threshold indicates ratio of tissue pixels}
		\label{patch}
	\end{center}
\end{figure*}

\subsection{Post-processing}
\begin{figure}[h]
	\begin{center}
		\includegraphics[scale=0.8]{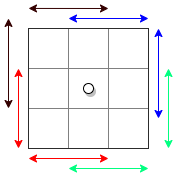}
		\caption{Each pixel was covered in four different patches i.e. each pixel will be predicted four times}
		\label{patch2}
	\end{center}
\end{figure}

\begin{figure}[h]
	\begin{center}
		\includegraphics[scale=0.4]{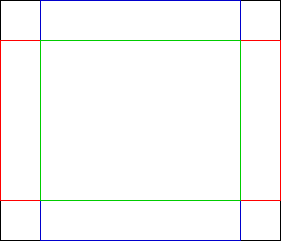}
		\caption{The green box is whole slide image. All other regions are padded portions. Patches  were taken from four different positions hence four different colors are used for depiction.}
		\label{patch1}
	\end{center}
\end{figure}

While predicting the given image, we predicted on patches of $512\times512\times3$ from the image with no overlapping and reconstructed the image. The predicted image, however, had block artifacts in it. Therefore, certain steps were taken as post-processing to overcome this and enhance the results. Firstly, the image of dimension $X\times Y$ was padded from all sides with a depth of 256, which resulted in the new dimensions $(X+512)\times(Y+512)$. Subsequently, the whole image was predicted upon with the above method by taking overlapping patches from 4 different starting points, (1) black:(0,0), (2) blue:(256,0), (3) red:(0,256) and (4) green:(256,256) as shown in figure \ref{patch2}. For representation, the center square is the original image, while the rest of the squares are padded regions. The four overlapping patches are then taken and predicted upon so that the original image is predicted upon four times. After this, the segmentation result corresponding to the original image in each of the four patches is extracted from the four predicted segmentation maps by removing the excess padding. The final result is then obtained by averaging the four segmentation maps and then thresholding for 50\%. The averaging provides a more robust output. A more accurate representation can be seen in figure \ref{patch1}, wherein the center green square is the original image while the rest of the squares and rectangles are padded regions. Also, the above technique of post-processing can be easily implemented in clinical settings as it only makes use of a simple padding algorithm.

\subsection{Loss Function}
The loss function used for training of model was the Dice Loss, which is the complement of the Dice Coefficient(DC). DC is the measure of the intersection or similarity between the two representations. It ranges from 0 to 1, where a DC of 1 denotes precise and whole overlap. The DC was formerly stated for binary data and calculated as:

\[DC = \frac{2*|X\cap Y|}{|X|+|Y|} \]
\[Loss = 1 - DC \]

where \(X\cap Y\) represents the common elements between sets \(X\) and \(Y\), and \(|X|\), \(|Y|\) represents the number of elements in set X and set Y, respectively.  \(|X\cap Y|\) was computed as the element-wise multiplication between the target and prediction mask and then adding the resulting matrix. The Dice Loss is a popular loss function used for various segmentation tasks, especially for medical image segmentation, and we used the same for training. The Dice Coefficient was used as the performance metric for comparison during testing.

\subsection{Ensemble Modeling}

The process of ensemble modeling remained reasonably straightforward. We used the self-ensemble \cite{liu2018towards} technique, where the results from six flipped/rotated versions of the same image were calculated. The final probability predictions were calculated by averaging all six predictions. We also ensemble the results obtained from our two best performing models, namely ResUNet and the proposed AtResUNet model.

\section{Experiments and Results}
We implemented our network using Keras (version 2.2.4) with Tensorflow backend. For preprocessing, OpenCV(version 4.1.0) and Scikit-Learn(version 0.21.2) was used. All the networks were trained on two NVIDIA Tesla-V100 GPUs with a mini-batch size of four. Adam \cite{adam} optimizer was used to optimize the whole network, and the learning rate was initialized as 0.001 and decayed, according to cosine annealing. The patch initially extracted from the dataset has the shape of $512\times512$, max-pooling was performed in subsequent layers till the resolution of the feature map reduced to 1/8th of the original. For better training and reduction in overfitting of the model, batch-normalization \cite{batch_norm}, and twenty-five percent dropout \cite{dropout} was applied before the max-pooling operation. Every convolution filter in the model is of size $3\times3$, and dilation rate, one unless specified. The activation function used is ReLU. In UNet \cite{uneto}, after every pooling operation, the output is passed through a $3\times3$ convolution, ReLU activation, and batch normalization layer. The model was trained for 100 epochs with data augmentation (as stated above) and a training dice coefficient of 0.96, and a validation dice coefficient of 0.87 was achieved. Lastly, even though the training was done using 30\% as the threshold for white matter ratio, the model performed exceptionally well on images with a ratio up to the tune of even 70\%.

\begin{table*}[]
\centering
	\begin{tabular}{|c|c|c|c|c|}
		\hline
		\textbf{Model / Architecture} &
		\textbf{Acc.}
		& \textbf{Dice} &  \textbf{Sen.} & \textbf{Spe.} \\ \hline
		FCN \cite{long2015fully}        & 0.473  & 0.497 & 0.461  & 0.502           \\ \hline
		DnCNN \cite{dncnn}       & 0.621    & 0.612 & 0.604 & 0.647               \\ \hline
		UNet\cite{uneto}               & 0.725         & 0.684 & 0.713
		& 0.737\\ \hline
		ResUNet \cite{isensee2020or}         & 0.734        & 0.725  & 0.741
		& 0.776\\ \hline
		Proposed AtResUNet    & 0.762      & 0.748 & 0.759 & 0.794            \\ \hline
		
		\textbf{Ensemble (AtResUNet+ResUNet)} & \textbf{0.788} & \textbf{0.753} & \textbf{0.767} & \textbf{0.802}
		
		\\ \hline
		
	\end{tabular}
	\caption{Comparison among the models evaluated on the basis of Accuracy(Acc.), Dice Coefficient, Sensitivity(Sen.), Specificity(Spe.).}
	\label{Results}
\end{table*}

\begin{figure*}[t]
	\begin{center}
		\includegraphics[width=\textwidth]{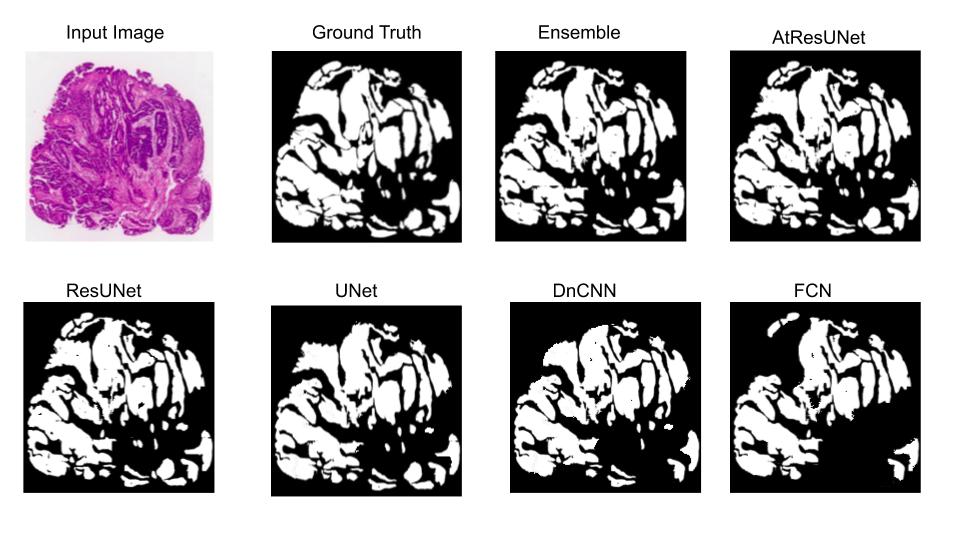}
		\caption{Qualitative Results of the final model}
		\label{results}
	\label{comp}
	\end{center}
	
\end{figure*}

Without limiting the study to UNet \cite{uneto} and its variants, we also trained other models as given in the works of  \cite{long2015fully}, and  \cite{dncnn}. As for the UNet-based models, we trained and compared the results with UNet itself and ResUNet \cite{isensee2020or}, which follows the idealogy of using residual connections in the UNet. For generalized results, self-ensembles of all the models were used for comparison. The winning team of the DigestPath 2019 Challenge achieved a dice score of 0.8075 on the testing dataset. However, we generated and compared our results using a five-fold cross-validation split of the training data itself. A considerable improvement from the UNet was noted with the use of residual connections, which was further improved upon using the novel Series Atrous Convolution Unit. It was found that the proposed model converges more effectively and can carry more information, which increased localization and segmentation accuracy of the model compared to that of UNet. Finally, the ensemble of the novel AtResUNet and the ResUNet was found to give the best results in our study as seen in table \ref{Results}, and figure \ref{comp}.

\section{Conclusion}
In this work, we have proposed a novel CNN-based architecture for segmenting the tumor in colonoscopy images. The AtResUNet combines atrous convolutions and residual connections with the UNet as the base model. Our architecture outperformed the existing architectures to emerge as the state-of-the-art for the task. Also, the pre and post-processing techniques used provided for efficient patch-based processing of the high-resolution images, which comprised a substantial amount of white matter. For overall comparison and generalization, the self-ensembling of all the architectures was done. Finally, the ensemble of the novel AtResUNet and ResUNet was found to give the best of the result in our study. Further work on this task includes introducing adversarial networks for the generation of artificial data for improved training. The adversarial network could also be directly introduced in the training process for better performance on the segmentation predictions. Lastly, we aim to make the model more generalizable to be used for other biomedical segmentation tasks too.

{\small
\bibliographystyle{splncs04}
\bibliography{egbib}
}

\end{document}